\documentstyle[multicol,aps,epsf]{revtex}

\begin{document}

\title{Charge Fluctuations in Quantum Point Contacts and 
Chaotic Cavities in the Presence of Transport}

\author{M.\ H.\ Pedersen$^a$, S.\ A.\ van Langen$^b$, and M.\ B\"uttiker$^a$}
\address{$^a$ D\'epartement de Physique Th\'eorique,
Universit\'e de Gen\`eve, 1211 Gen\`eve 4, Switzerland}
\address{$^b$ Instituut-Lorentz, Leiden University, P.O.\ Box 9506, 2300 RA Leiden, 
The Netherlands}

\date{\today}
\maketitle

\begin{abstract}
We analyze the frequency-dependent current fluctuations induced into a gate
near a quantum point contact or a quantum chaotic cavity. 
We use a current and charge conserving, effective scattering approach 
in which interactions are treated in random phase approximation.
The current fluctuations measured at a nearby gate, coupled capacitively
to the conductor, are determined by the 
screened charge fluctuations of the conductor.
Both the equilibrium and the non-equilibrium current noise at the gate
can be expressed with the help of resistances 
which are related to the charge dynamics on the conductor. 
We evaluate these resistances for a point contact and 
determine their distributions for an ensemble of chaotic 
cavities. For a quantum point contact these resistances exhibit pronounced
oscillations with the opening of new channels. 
For a chaotic cavity coupled to one channel point contacts 
the charge relaxation resistance shows a broad distribution between 
1/4 and 1/2 of a resistance quantum. The non-equilibrium resistance 
exhibits a broad distribution between zero and 1/4 of a resistance 
quantum. 

\end{abstract}


\begin{multicols}{2}

\section{Introduction}

The investigation of fluctuations in mesoscopic conductors
is an interesting problem which has found considerable 
attention both experimentally and theoretically. Two recent reviews 
provide both an introduction to the subject as well as a discussion
of some of the important results \cite{dejong,buttik96}.
In this work we are interested in the frequency-dependent noise 
spectra of mesoscopic conductors away from the low-frequency 
white-noise limit. The experimental observation of deviations from 
the white noise-limit in the current-fluctuation 
spectra of well conducting samples requires large frequencies\cite{prober97}. 
Here we investigate the fluctuations induced into a nearby gate, capacitively 
coupled to the conductor. 
These fluctuations are 
not a correction to an effect that exists already in the zero-frequency
limit. 
We present a discussion which describes the 
internal potential of the mesoscopic conductor with 
a single variable. The Coulomb interactions are described 
with the help of a geometrical capacitance $C$ instead of the 
full Poisson equation. 
Furthermore, we will treat the gate as a macroscopic
electric conductor. In this case the current fluctuations induced into
a nearby gate are determined entirely by the dynamics of the charge 
fluctuations of the mesoscopic conductor. 

\begin{figure}
\narrowtext
\vspace*{0.5cm}
\epsfysize=7cm
\epsfxsize=7cm
\centerline{\epsffile{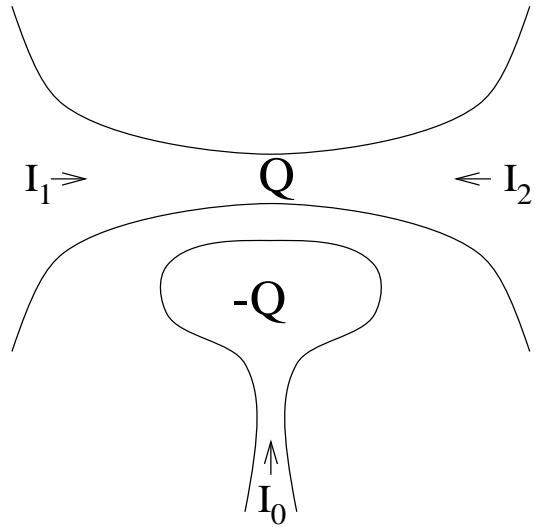}}
\vspace*{0.5cm}
\caption{ \label{qpc_geometry}
Geometry of the quantum point contact.
}
\end{figure}

Consider a conductor, for instance the 
quantum point contact\cite{vanwees,wharam,reznikov,kumar},
shown in Fig.~\ref{qpc_geometry}.
The conductor is  
described by scattering matrices 
$s_{\alpha\beta}$ which relate the amplitudes of incoming currents at 
contact $\beta$ to the amplitudes of the outgoing currents at 
$\alpha$. We find that the charge fluctuations of the mesoscopic conductor
can be described with the help of a density of states matrix 
\begin{equation}
{\cal N}_{\delta\gamma} = \frac{1}{2\pi i} \sum_\alpha 
s_{\alpha\delta}^\dagger \frac{ds_{\alpha\gamma}}{dE}.
\label{d0}
\end{equation}
The diagonal elements of this matrix determine the density of states
of the conductor $N = \sum_{\gamma} \mbox{Tr} ({\cal N}_{\gamma\gamma})$;
the trace is over all quantum channels. 
The non-diagonal elements are essential to describe fluctuations. 
At equilibrium and in the zero temperature limit we find 
that to leading order in frequency the mean squared 
current fluctuations at the gate have a spectrum 
$S_{00}(\omega,V=0) = 2\omega^2 \hbar|\omega| C_\mu^2 R_q .$
Here 
$C_\mu^{-1} = C^{-1} + ( e^2 N )^{-1}$,
is the {\em electro-chemical} capacitance\cite{btp93}
of the conductor vis-{\`a}-vis the gate. 
The dynamical quantity which determines 
the fluctuations is the charge relaxation resistance $R_q$
\begin{eqnarray}
    R_q &=& \frac{h}{2e^2} \frac{\sum_{\gamma\delta} \mbox{Tr} 
    \left( \cal{N}_{\gamma\delta} \cal{N}^\dagger_{\gamma\delta} \right)}
	{[\sum_{\gamma} \mbox{Tr}({\cal N}_{\gamma\gamma})]^{2}} .
	\label{Rq} 
\end{eqnarray}
B\"uttiker, Thomas and Pr\^etre\cite{btp93} showed that the charge 
relaxation resistance governs the dissipative part of the low frequency 
admittance of mesoscopic capacitors. 
Together with the electrochemical capacitance $C_\mu$, $R_q$
determines the charge relaxation time $R_qC_\mu$
of the mesoscopic conductor.
Similarly to the equilibrium noise spectrum, at zero
temperature, the non-equilibrium current noise 
spectrum at the gate, 
$S_{00}(V, \omega) = 2\omega^2 e|V| C_\mu^2 R_v$,
is determined by a resistance $R_v$,
\begin{eqnarray}
R_v  = \frac{h}{e^2} \frac{\mbox{Tr} 
    \left( {\cal N}_{21} {\cal N}^\dagger_{21}\right)}
    {[\sum_{\gamma} \mbox{Tr}({\cal N}_{\gamma\gamma})]^{2}}.
	\label{Rv} 
\end{eqnarray}
Whereas the charge relaxation resistance $R_q$ invokes all 
elements of the density of states matrix with equal
weight, in the presence of transport the non-diagonal elements 
of the density of states matrix are singled out. 
Below we present the derivation of these results and evaluate 
the charge relaxation resistance $R_q$ and the resistance $R_v$ 
for the quantum point contact and for a 
chaotic quantum dot.

The characterization of the current fluctuations 
in terms of resistances can be motivated as follows.
The current fluctuations at the gate contact are directly 
related to fluctuations of the charge $Q$ on the conductor, 
\begin{eqnarray}	
S_{00}(\omega, V) = \omega^{2} S_{QQ}(\omega, V).
\label{siq}
\end{eqnarray}
In turn, the charge fluctuations are related to the potential 
fluctuations by the geometrical capacitance $C$
\begin{eqnarray}
   S_{QQ}(\omega, V) = C^{2} S_{UU}(\omega, V).
\label{squ}
\end{eqnarray}
Voltage fluctuations, as is well known, are essentially determined by 
resistances. However, in contrast to the Nyquist formula for equilibrium
voltage fluctuations, we deal here with electro-static potential 
fluctuations inside the conductor. The resistances 
$R_q$ and $R_v$ are related to the charge dynamics rather than 
the two terminal dc-resistance.

The resistances $R_q$ and $R_v$ probe an aspect of mesoscopic conductors
which is not accessible by investigating the dc-conductance or the 
zero-frequency limit of shot noise. These resistances are not determined 
by the scattering matrix alone but also by its energy derivative.   
According to the fluctuation
dissipation theorem,
the low-frequency equilibrium current-fluctuations of a conductor 
which permits transmission are determined by the conductance of the system. 
For a two-terminal conductor the conductance is simply the 
sum of all transmission eigenvalues $T_n$. The low-frequency
non-equilibrium noise, the shot noise\cite{khlus,lesovik}, of
a two-terminal conductor is determined by the sum of the products 
$T_n (1-T_n)$, where the $T_n = 1-R_n$ are again the eigenvalues of 
the transmission matrix multiplied by its hermitian 
conjugate\cite{buttiker90,martin}. 
Hence both the equilibrium noise and the shot noise are governed 
by the transmission behavior of the sample. This is even true 
for correlations on multiterminal conductors which cannot be expressed in 
terms of transmission eigenvalues\cite{buttiker90,blanter97,stijn97}. 
In contrast, the dynamic conductance is determined by 
oscillations of the charge distribution in the conductor \cite{buttiker92_2}.
Since charge is a conserved quantity, the oscillatory part of the 
charge distribution can be represented as a sum of 
dipoles\cite{christen96,buchr96}.
Similarly, the frequency-dependent fluctuations are governed 
by the fluctuations of the charge distributions or more precisely
by the fluctuations of dipolar charges. 

The charge-fluctuations of a non-interacting system can be described 
with the help of the density of states matrix
Eq.\ (\ref{d0}). However, the charge distribution of 
a non-interacting system is not dipolar. In fact without 
interactions, charge is not conserved and consequently currents are not
conserved. To achieve a dipolar (or higher order multipolar) charge 
distribution it is necessary to consider interactions. 
Here we consider the simple approximation in which the charge 
distribution is effectively represented by a single dipole. 
We permit the charging of the quantum point contact vis-{\`a}-vis the
gate. In Fig.~\ref{qpc_geometry} this dipole is indicated by the charges 
$Q$ and $-Q$. A more realistic treatment of the charge distribution
of a quantum point contact includes a dipole across the quantum point 
contact itself\cite{christen96} and in the presence of the gates 
includes a quadrupolar charge distribution\cite{buchr96}.

The frequency-dependence of the noise spectra generated by the fluctuations 
of the dipolar charges should be distinguished 
from a purely statistical frequency-dependence arising  
from the Fermi distribution functions\cite{prober97,yang,ueda93}: 
Even for a conductor with an energy-independent
scattering matrix there exists a frequency-dependence due 
the Fermi distribution functions 
of the different reservoirs. For small frequencies the distribution 
functions are governed by the temperature $kT$ or the applied voltage $eV$
and a crossover occurs when the frequency $\hbar \omega$
exceeds both $kT$ and $eV$. We will not further emphasize this crossover 
since it is a property of the Fermi distribution alone and provides
no new information on the conductor itself. 

Our work is also of interest in view of recent efforts to 
discuss the dephasing induced by the shot noise of two
conductors in close proximity\cite{aleiner,levinson} or due to the
fluctuating electro-magnetic field \cite{webb}.
Our work shows that what counts are 
the dipolar charge fluctuations. 
The discussion presented below cannot be applied to metallic diffusive 
conductors for which the potential needs to be treated as a 
field\cite{buttik96}.
Recently Nagaev's\cite{nagaev93} classical discussion of shot noise 
in metallic conductors has been extended to investigate the effect of a 
nearby gate\cite{nagaev97,naveh97}. In these works the source of the noise
is taken to be 
frequency independent 
over the entire range of interest. 
In contrast for the examples treated here it is not only the 
electrodynamic response which is frequency dependent but also the noise itself. 

There has been a considerable recent interest in 
the parametric derivatives of the scattering matrix of chaotic
conductors \cite{fyod2,gopar,brouwer1,fyod,brouwer2}.
The energy derivative of the scattering 
matrix determines quantities
like the density of states matrix Eq.\ (\ref{d0}).
For electrostatic problems it is the functional derivative 
of the scattering matrix with respect 
to the local potential which matters \cite{but93}.  
Only in the limit where we describe the internal electrostatic potential 
as a single variable (instead of a continuous field), and only 
if we are satisfied with a WKB like-description, can the energy 
derivatives of the scattering matrix be used.
These two conditions are likely to be fullfilled for a ballistic quantum dot.
Then the
energy derivative of the scattering 
matrix and the potential derivative 
differ just by a sign. 
For chaotic cavities a theory of the energy derivative of the 
scattering matrix has permitted a 
discussion of the distribution of capacitances \cite{gopar,brouwer1,fyod}.
Fyodorov and Sommers\cite{fyod2,fyod} used supersymmetric 
methods to investigate the energy derivatives of the scattering
matrix. For a chaotic cavity connected to a reservoir via a single channel
lead, 
Gopar, Mello and B\"uttiker\cite{gopar} 
found the distribution functions for all universality
classes by analyzing directly the statistical properties 
of the scattering matrix. The single channel discussion of 
Gopar et al.\cite{gopar} has been generalized by  
Brouwer, Frahm and Beenakker\cite{brouwer2}, who
found the 
distribution of the scattering matrix and its derivatives
for the multi-channel problem.
This generalization made it possible to investigate the distribution 
of parametric conductance derivatives like the 
transconductance $dG/dV_0$, where $G$ is the conductance and 
$V_0$ is the gate voltage\cite{brouwer3}. 
Here we use the result of Ref.\ \cite{brouwer2}
to find the distribution of the charge relaxation resistance
$R_q$ and the resistance $R_v$ for a chaotic quantum dot coupled
to reservoirs via two perfect one-channel leads. 

\section{Current and Charge Fluctuations} 

To find the current-fluctuations 
for the structures of interest 
we discuss in this section an approach which 
includes interaction effects in the random phase approximation (RPA). 
This approach has been used in Ref.~\cite{bpt93}
to find the dynamic conductance of mesoscopic structures for the 
case that the self-consistent potential of the 
conductor can be taken to be a single variable $U$.
The fluctuations belonging to this approach are discussed in
Ref.~\cite{btp93} for the case of a mesoscopic capacitor 
and for a more general multiprobe conductor capacitively coupled 
to a gate in Ref.~\cite{stlouis}. 

\subsection{Fixed Internal Potential}

We consider a conductor with a fixed internal 
potential (non-interacting problem) and present the results 
needed later on to 
treat the problem with interactions. 
Consider a conductor described by scattering matrices 
$s_{\alpha\beta}$ which relate the 
annihilation operators $\hat{a}_\beta$
in the incoming channels in contact $\beta$ 
to the annihilation operators $\hat{b}_\alpha$ of a carrier in the 
outgoing channel of contact $\alpha$
via\cite{buttiker90}
\begin{equation}
    \hat{b}_\alpha = \sum_\beta s_{\alpha\beta} \hat{a}_\beta .
\label{smatrix}    
\end{equation}
In a multichannel conductor the $s$-matrix 
has dimensions $N_{\alpha}\times N_{\beta}$ 
for leads that support $N_{\alpha}$ and $N_\beta$ quantum channels.
Here $\alpha$ and $\beta$ run over all contacts of the
conductor $\alpha, \beta = 1 ,2$. (Later, we need 
indices for the contacts of the conductor and the gate. 
For this case we will use the labels $\mu, \nu = 0, 1, 2$ ).  
The current at contact $\alpha$ is determined by the difference 
in the occupation of the incident channels minus the occupation
of the outgoing channels 
\begin{equation}
\hat{I}_\alpha(\omega) = \frac{e}{\hbar} \int dE [
\hat{a}^{\dagger}_{\alpha} (E) \hat{a}_{\alpha} (E+\hbar\omega) -
\hat{b}^{\dagger}_{\alpha} (E) \hat{b}_{\alpha} (E+\hbar\omega) ].
\label{cur1}
\end{equation}
Using Eq.\ (\ref{smatrix}) to eliminate the occupation numbers of the 
outgoing channels in terms of the incoming channels 
yields a current operator\cite{buttiker90} 
\begin{equation}
\hat{I}_\alpha(\omega) = \frac{e}{\hbar} \int dE \sum_{\beta\gamma}
\hat{a}_\beta^\dagger(E) A^0_{\beta\gamma}(\alpha,E,E+\hbar\omega)
\hat{a}_\gamma(E+\hbar\omega),
\label{curo}
\end{equation}
with a {\em current matrix} 
\begin{equation}
A^0_{\delta\gamma}(\alpha,E,E^{\prime}) = \delta_{\alpha\delta}
\delta_{\alpha\gamma} 1_\alpha - 
s_{\alpha\delta}^\dagger(E) s_{\alpha\gamma}(E^{\prime}).
\label{curm}
\end{equation}
Here the upper index $0$ indicates that we 
deal with non-interacting electrons.
The current noise spectra are determined by the quantum expectation
value $\langle\cdots\rangle$ of the current operators at 
contact $\mu$ and $\nu$, 
$\frac{1}{2}\langle \Delta \hat{I}_\mu(\omega) \Delta \hat{I}_\nu(\omega') 
+\Delta \hat{I}_\nu(\omega') \Delta \hat{I}_\mu(\omega) 
\rangle \equiv 2 \pi S_{\mu\nu} \delta(\omega+\omega')$. 
The spectral densities in terms of the current matrix are\cite{buttiker90}
\begin{eqnarray}
    S_{\mu\nu}(\omega) &=& \frac{e^2}{h} \sum_{\delta\gamma} \int dE\,
	F_{\gamma\delta}(E,\omega) \label{s0} \\
	&\mbox{Tr}& [ A^0_{\gamma\delta}(\mu,E,E+\hbar\omega) 
	(A^0)_{\gamma\delta}^\dagger (\nu,E,E+\hbar\omega) ] , \nonumber \\
    F_{\gamma\delta}(E,\omega) &=& f_\gamma(E)(1-f_\delta(E+\hbar\omega))\nonumber \\
	&& + f_\delta(E+\hbar\omega)(1-f_\gamma(E)). \label{fermi} 
\end{eqnarray}
Here the trace is taken over channels and $f_\gamma$ is the Fermi distribution function
for contact $\gamma$.
At equilibrium these fluctuation spectra are related to the ac-conductances of 
the non-interacting problem discussed in Ref.~\cite{bpt93}.
The current operator for the gate has thus far not been defined: that will 
be achieved only in the next section.

It is natural to decompose the current matrix into two contributions,
one at equal energies determines the dc-response of the conductor
and one at differing energies is associated with 
the dynamics of the system.
Thus we write  
\begin{eqnarray}
    A^0_{\delta\gamma}(\alpha,E,E^{\prime}) &=& \delta_{\alpha\delta}
    \delta_{\alpha\gamma} 1_\alpha - 
    s_{\alpha\delta}^\dagger(E) s_{\alpha\gamma}(E) \nonumber \\ 
    && -i (E^{\prime} - E) {\cal N}_{\delta\gamma}(\alpha, E,E^{\prime})
\label{ad}
\end{eqnarray} 
with a {\em partial density of states matrix} 
\begin{equation}
{\cal N}_{\beta\gamma}(\alpha, E,E') = 
\frac{i}{2\pi} \frac{s_{\alpha\beta}^\dagger(E) \left( s_{\alpha\gamma}(E)-
s_{\alpha\gamma}(E')\right) }{E'-E}. 
\label{density_matrix}
\end{equation}
This matrix has a simple interpretation: The elements of 
${\cal N}_{\beta\gamma}(\alpha, E,E')$ are 
the diagonal and non-diagonal elements of the density of 
states associated with carriers incident from contact $\beta$ and 
$\gamma$ which eventually contribute to the 
current at contact $\alpha$.
From the continuity equation we find  
immediately that the total charge fluctuations in the 
conductor generated by particles incident 
from contact $\beta$ and $\gamma$ 
irrespective through which contact 
they leave the conductor are determined by the density of states matrix 
\begin{equation}
{\cal N}_{\beta\gamma}(E,E') = 
\sum_\alpha {\cal N}_{\beta\gamma}(\alpha, E,E').
\label{dmatrix}
\end{equation} 
Some additional combinations of these matrices have a special meaning 
\cite{but93,gasparian}.
We call 
\begin{equation}
   \overline{\cal N}_{\beta}(E,E') = \sum_\alpha {\cal N}_{\beta\beta}(\alpha, E,E')
\label{inmatrix}
\end{equation} 
the {\em injectance matrix} of contact $\beta$ and call
\begin{equation}
\underline{\cal N}_{\alpha}(E,E') = 
\sum_\beta{\cal N}_{\beta\beta}(\alpha, E,E')
\label{emmatrix}
\end{equation} 
the {\em emittance matrix}. 
The frequency dependent injectance is 
the quantum expectation value of the injectance operator 
$\overline{N}_{\beta}(\omega) = 
\langle \sum_\alpha \hat{\cal N}_{\beta\beta}(\alpha, E,E+\hbar\omega)\rangle$.
Similarly the frequency dependent emittance is 
$\underline{N}_{\alpha}(\omega) =
\langle\sum_\beta \hat{\cal N}_{\beta\beta}(\alpha, E,E+\hbar\omega)\rangle$.
Below we will often use only the zero-frequency limit 
of the density matrix Eq.~(\ref{dmatrix}) ($\omega \rightarrow 0$)
which is given by Eq.~(\ref{d0}).
Similarly we will most often use only the zero-temperature, 
zero-frequency injectance, 
\begin{equation}
    \overline{N}_{\beta} = \frac{1}{2\pi i} \sum_\alpha 
    \mbox{Tr} \left( s_{\alpha\beta}^\dagger \frac{ds_{\alpha\beta}}{dE} \right)
\label{i0}
\end{equation}
and emittance,
\begin{equation}
    \underline{N}_{\alpha}
    = \frac{1}{2\pi i} \sum_\beta 
    \mbox{Tr} \left( s_{\alpha\beta}^\dagger \frac{ds_{\alpha\beta}}{dE} \right).
\label{e0}
\end{equation}
The density matrices introduced above together with 
the injectances and emittances 
can now be used to characterize the 
charge fluctuations of the conductor. 
The evaluation of the injectances and emittances in the 
equilibrium state of the conductance limits the theory
presented below to linear order in the applied voltage.

\subsection{Effective Current Matrix}

Our goal is to derive a current matrix which includes 
the effect of screening \cite{buttik96} and replaces the current 
matrix, Eq.~(\ref{curm}) of the non-interacting problem.
To this extent we next determine the operator $\hat U$ for the
internal potential. The charge on the conductor is 
determined by the Coulomb interaction. 
Here we describe the interaction with the help of a 
single geometrical capacitance. Hence the charge on the 
conductor is $\hat Q = C \hat U$. Here we have assumed that the gate 
is macroscopic and has no dynamics of its own. 
We can also determine the charge $\hat Q$ 
as the sum of
the bare charge fluctuations $e \hat {\cal N}$ and 
the induced charges generated by the fluctuating induced 
electrical potential. In RPA the induced charges are 
proportional to the average frequency dependent density of states 
$N(\omega)$ times the fluctuating potential.
Thus the net charge is determined by 
\begin{equation}
\hat Q = C \hat U = e\hat{\cal N} - e^{2}N\hat U.
\end{equation}

Solving this equation gives us for the operator of the 
potential fluctuations 
\begin{equation}
\hat U = G e \hat{\cal N},
\end{equation}
with 
\begin{equation}
    G (\omega) = (C + e^{2} N(\omega))^{-1} .
\end{equation}
Here $G(\omega)$ takes into account the 
effective interaction potential.   

The total current at probe $\alpha$ is determined by the 
particle current and in addition by a current 
due to the fluctuating potential. The fluctuation of the 
internal potential creates additional currents
at all the contacts. The current fluctuations 
generated by the induced potential 
fluctuations at contact $\alpha$ are determined by
$i\omega e^2 \underline{N}_\alpha(\omega) \hat{U} (\omega)$.
Here the response to the internal potential is 
determined by the emittance\cite{but93,bpt93} of the conductor
into contact $\alpha$.
Thus the total current at contact $\alpha$ of the conductor is 
\begin{equation}
\hat{I}_\alpha(\omega) = \hat{I}_\alpha^0(\omega) - 
i\omega e^2 \underline{N}_\alpha(\omega) \hat{U} (\omega) ,
\label{curc}
\end{equation}  
where $\hat{I}_\alpha^0$ is the current operator for fixed internal
potential.
The current induced into the gate is given by the time derivative of the total 
charge and hence by 
\begin{equation}
\hat{I}_g(\omega)= i \omega C \hat{U} (\omega) .
\label{curg}
\end{equation}
Expressing $\hat{U}$ in terms of the density of states matrix gives for 
the current operators Eqs.~(\ref{curc}) and (\ref{curg}) 
an expression which is of the same form as Eq.~(\ref{curo}) but 
with the current matrix Eq.~(\ref{curm}) replaced by an 
effective current matrix  
\begin{eqnarray} 
    A_{\delta\gamma}(\alpha,E,E+\hbar\omega) &=& 
        A^0_{\delta\gamma}(\alpha,E,E+\hbar\omega) \nonumber \\
	&& + i\omega e^2 \underline{N}_\alpha 
	G {\cal N}_{\delta\gamma}(E,E+\hbar \omega).
	\label{aeffc}
\end{eqnarray}
Eq.~(\ref{aeffc})	
determines the current at the contacts of the conductor. 
The current induced into the gate contact is determined by
a current matrix
\begin{eqnarray}	
    A_{\delta\gamma}(0,E,E+\hbar\omega) = 
	- i\omega C G {\cal N}_{\delta\gamma}(E,E+\hbar\omega).
	\label{aeffg}
\end{eqnarray}
The sum of all currents at the contacts of the sample and the current 
at the gate is conserved. Indeed, labelling the index which runs 
over all contacts by $\nu$, $(\nu = 0, 1, 2)$ we find
\begin{eqnarray}	
    \sum_\nu A_{\delta\gamma}(\nu,E,E+\hbar\omega) = 0.
    \label{suma}
\end{eqnarray}
Eq.~(\ref{suma}) follows from the relation between the 
bare current matrix and the density 
of states matrix, Eqs.~(\ref{ad}-\ref{dmatrix})
and the fact that $1-e^{2}NG = CG$.   
Before continuing we notice that for these effective current matrices
$A_{\delta\gamma}(\nu)$, the index $\nu$ runs over all contacts
but the indices $\delta$ and $\gamma$ run only over the contacts 
of the sample. This ``asymmetry'' is a consequence of our macroscopic 
treatment of the gate.

\subsection{Charge Fluctuation Spectra}

With the help of the effective current matrices Eqs.~(\ref{aeffc})
and (\ref{aeffg}) we can find the current fluctuation spectra 
$S_{\mu\nu}(\omega, V)$ 
as in the non-interacting case: In Eq.~(\ref{s0}) we have to replace 
the bare current matrix $A^0_{\delta\gamma}(\alpha)$ by the effective 
current matrix $A_{\delta\gamma}(\nu)$. 
This determines a matrix $S_{\mu\nu}(\omega)$ of fluctuation spectra for the 
mean square current fluctuations at the contacts of the conductor 
and the gate and for the correlations between any two currents. 
As a consequence of current conservation  
$\sum_{\mu}S_{\mu\nu}(\omega) = \sum_{\nu}S_{\mu\nu} (\omega) = 0.$ 
At equilibrium the fluctuation spectra which we find with the help of the
effective current matrix are related via the fluctuation dissipation
theorem to the frequency dependent conductances 
of the interacting system given in Ref.\ \cite{bpt93}.
The spectra also agree with the expression given in Ref.\ \cite{stlouis}. 
Here we are interested in the current fluctuations at the gate
determined by the spectrum $S_{00}(\omega, V)$.
This spectrum is entirely determined by the charge fluctuations
of the conductor (see Eq.~(\ref{siq})). 
Defining the frequency dependent capacitance of the conductor to the gate 
$C_{\mu} (\omega) \equiv e^{2}N(\omega)CG(\omega)$ 
and using Eq.~(\ref{aeffg}) we find  
\begin{eqnarray}
    S_{QQ}(\omega) &=& C_{\mu}^{2}(\omega) N^{-2}(\omega)
	\sum_{\delta\gamma} \int dE\, F_{\gamma\delta}(E,\omega) \nonumber\\
	&& \mbox{Tr} [{\cal{N}}_{\gamma\delta}(E,E+\hbar\omega) 
	{\cal{N}}_{\gamma\delta}^\dagger (E,E+\hbar\omega) ].
\label{qfluct} 
\end{eqnarray}
Two limits are of special interest.
At equilibrium, at zero temperature, we find for the 
charge fluctuation spectrum in the low frequency limit, 
$S_{QQ}(\omega) = 
2 C_{\mu}^{2} R_{q} \hbar |\omega|$ where 
the electro-chemical capacitance is given by its 
zero-frequency value and where  
the charge relaxation resistance is determined by Eq.~(\ref{Rq}).


The second limit we wish to consider is the zero-temperature, 
low-frequency limit of the charge fluctuations 
to leading order in the applied
voltage $V$. Evaluation of Eq.~(\ref{qfluct}) 
gives $S_{QQ}(\omega) = 2 C_{\mu}^{2} R_{v} |eV|$ with a resistance 
$R_v$ given by Eq.~(\ref{Rv}). 
Thus the non-equilibrium noise is determined by a non-diagonal
element of the density of states matrix. If both the frequency and 
the voltage are non-vanishing we obtain to leading order
in $\hbar \omega$ and $V$, $S_{QQ}(\omega) = 
2 C_{\mu}^{2} R(\omega,V) \hbar |\omega|$ with a resistance 
\begin{equation}
    R(\omega,V)\hbar|\omega|=\left\{
	\begin{array}{ll}
	    R_q \hbar|\omega| , & \hbar|\omega| \geq e|V| \\
	    R_q \hbar|\omega| + R_V (e|V|-\hbar|\omega|) , &
		\hbar|\omega| \leq e|V|
\end{array} \right. \label{effectiveR}
\end{equation}
which is a frequency and voltage 
dependent series combination of the resistances $R_{q}$ and $R_{v}$. 
Below we discuss the resistances $R_{q}$ and $R_{v}$ in detail for 
two examples: a quantum point contact and a chaotic cavity. 

\section{Quantum point contact} \label{qpc}

Quantum point contacts are formed with the help of gates. 
It is therefore interesting to ask what the fluctuations
are which would be measured at one of these gates.
For simplicity, we consider a
symmetric contact: We assume that the electrostatic potential is 
symmetric for electrons approaching the contact from the left or from 
the right. 
Furthermore we combine the capacitances 
of the conduction channel to the two gates and consider 
a single gate as schematically shown in  Fig.~\ref{qpc_geometry}.
If only a few channels are open the potential 
has in the center of the conduction channel 
the form of a saddle\cite{butqpc}:
\begin{equation}
    V(x,y) = V_0 + \frac{1}{2} m \omega_y^2 y^2
    - \frac{1}{2} m \omega_x^2 x^2
\end{equation}
where $V_0$ is the electrostatic potential at the saddle and the curvatures of
the potential are parametrized by $\omega_x$ and $\omega_y$.
For this model the scattering matrix is diagonal, 
i.e.\ for each quantum channel ( energy $\hbar \omega_y (n+1/2)$
for transverse motion)
it can be represented as a $2\times 2$-matrix.
For a symmetric scattering potential and without a magnetic field
the scattering matrix is of the form
\begin{equation}
    s_{n}(E) = \left( \begin{array}{ll}
	-i \sqrt{R_{n}} \exp(i\phi_{n}) & \sqrt{T_{n}} \exp(i\phi_{n}) \\
	\sqrt{T_{n}} \exp(i\phi_{n}) & -i\sqrt{R_{n}} \exp(i\phi_{n})
	\end{array} \right)
\end{equation} 
where $T_{n}$ and $R_{n}= 1-T_{n}$
are the transmission and reflection probabilities 
of the n-th quantum channel and $\phi_{n}$ is the phase accumulated by a carrier
in the n-th channel
during transmission through the QPC. 
The probabilities for transmission through the saddle point are\cite{butqpc} 
\begin{eqnarray}
    T_{n}(E) &=& \frac{1}{1+e^{-\pi\epsilon_n(E)}} ,\\
    \epsilon_n(E) &=& 2\left[ E-\hbar\omega_y(n+\frac{1}{2})-V_0\right]/
	(\hbar\omega_x) .
\end{eqnarray}
The transmission probabilities determine the conductance 
$G = ({e^2}/{h}) \sum_{n} T_{n}$ and the zero-frequency
shot-noise \cite{lesovik,buttiker90}
$S(\omega = 0, V) = ({e^2}/{h}) (\sum_{n} T_{n} R_{n}) e|V|$. 
As a function of energy (gate voltage) the conductance rises 
step-like\cite{vanwees,wharam}. The shot noise 
is a periodic function of energy. The oscillations 
in the shot noise associated with the opening of a quantum channel 
have recently been demonstrated experimentally
by Reznikov et al.\cite{reznikov} and Kumar et al.\cite{kumar}.   

\begin{figure}
\narrowtext
\epsfysize=9cm
\epsfxsize=7cm
\centerline{\epsffile{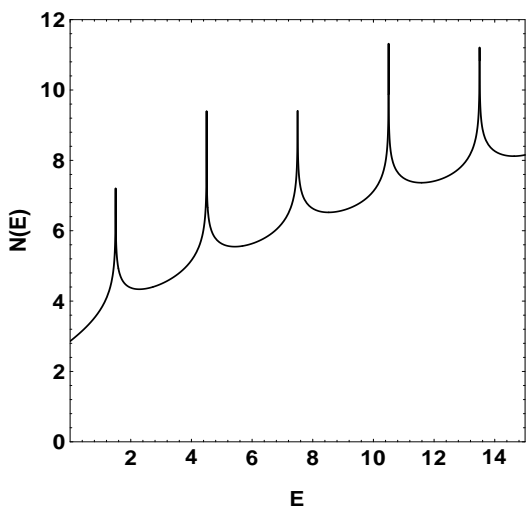}}
\caption{ \label{dos_fig}
Density of states in units of $4/(h\omega_x)$ for a saddle-point constriction 
as function of energy, $E/(\hbar\omega_x)$.
}
\end{figure}

To obtain the density of states we use the relation between 
density and phase $N_{n} = (1/\pi) \phi_{n}$
and evaluate it semi-classically.
The spatial region of interest for which we have to find 
the density of states is the region over which the electron
density in the contact is not screened completely. 
We denote this length by $\lambda$. 
The density of states is then found from  
$N_{n}=1/h \int_{-\lambda}^{\lambda} \frac{dp_n}{dE} dx$ 
where $p_{n}$ is the classically allowed momentum.
A simple calculation gives a density of states 
\begin{equation}
    N_{n} (E) = \frac{4}{h\omega_x}
    \mbox{asinh} \left( \sqrt{\frac{1}{2} \frac{m\omega_x^2}{E-E_n}}\lambda  \right),
\end{equation}
for energies $E$ exceeding the channel threshold $E_n$ and 
\begin{equation}
   N_{n} (E) = \frac{4}{h\omega_x}	    
   \mbox{acosh} \left( \sqrt{\frac{1}{2} \frac{m\omega_x^2}{E_n-E}}\lambda \right) ,
\end{equation}
for energies in the 
interval $E_n - (1/2) m \omega_x^2 \lambda^2 \leq E < E_n$ 
below the channel threshold. 
Electrons with energies less than $E_n - \frac{1}{2} m \omega_x^2 \lambda^2 $
are reflected before
reaching the region of interest, and thus do not contribute to the density of states.
The resulting density of states has a logarithmic singularity
at the threshold $E_{n}= \hbar\omega_y(n+\frac{1}{2})+V_0$  
of the n-th quantum channel. (We expect that a fully quantum mechanical 
calculation gives a density of states which exhibits also
a peak at the threshold but which is not singular). 
The total density of states as function 
of energy (gate voltage) is shown in Fig.~\ref{dos_fig} 
for 
$\omega_y/\omega_x=3$, $V_0=0$ and $m\omega_x \lambda^2/\hbar=18$.
Each peak in the density of states of Fig.~\ref{dos_fig} 
marks the opening of a new channel. With the help of 
the density of states we also obtain the capacitance 
$C_\mu^{-1} = C^{-1} + ( e^2 N )^{-1}$. For the experimentally 
most relevant case $(e^{2}/C) \gg N^{-1}$ the variations 
in the capacitance are small and the noise spectra 
are dominated by the energy dependence of $R_q$ and $R_v$ 
which we will now discuss. 

\begin{figure}
\narrowtext
\epsfysize=9cm
\epsfxsize=7cm
\centerline{\epsffile{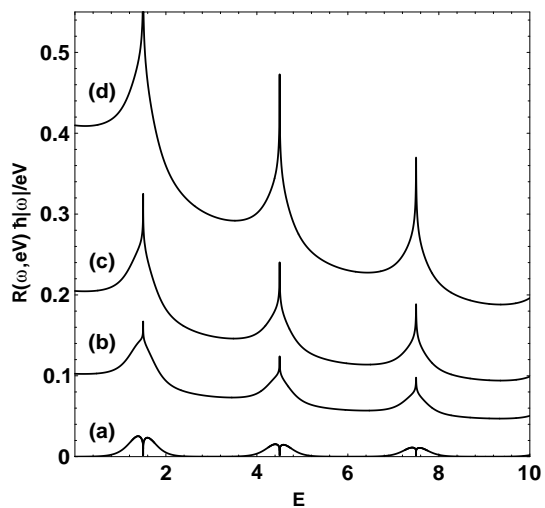}}
\caption{ \label{R_fig}
Effective resistance, in units of $h/e^2$, as function of
energy, $E/(\hbar\omega_x)$ for the cases
$a)\hspace{0.15cm} \hbar\omega/(eV)=0$, 
$b)\hspace{0.15cm} \hbar\omega/(eV)=0.25$, 
$c)\hspace{0.15cm} \hbar\omega/(eV)=0.5$ and 
$d)\hspace{0.15cm} \hbar\omega/(eV)=1$, 
where $V$ is bias voltage.}
\end{figure}

It is instructive to evaluate the resistances 
$R_q$ and $R_v$ explicitly in terms of the parameters 
which determine the scattering matrix. 
We find for the density of states matrix 
of the n-th quantum channel
\begin{eqnarray}
    {\cal N}_{11} &=& {\cal N}_{22} = \frac{1}{2\pi} \frac{d\phi_{n}}{dE},\\
    {\cal N}_{12} &=& {\cal N}_{21} = \frac{1}{4\pi} 
    \frac{1}{\sqrt{R_{n}T_{n}}} \frac{dT_{n}}{dE}.
\end{eqnarray}
Inserting these results into Eq.~(\ref{Rq}) 
gives for the charge relaxation resistance\cite{bpt93} 
\begin{eqnarray}
    R_q &=& \frac{h}{e^2} \frac{\sum_{n} (d\phi_{n}/dE)^{2}}
    {[\sum_{n} (d\phi_{n}/dE)]^{2}} .
\end{eqnarray}
It is determined by the derivatives 
of the phases (densities) evaluated at the Fermi energy. 
The resistance $R_v$ is given by 
\begin{eqnarray}
R_v &=& \frac{h}{e^2} \frac{ \sum_n \frac{1}{4R_nT_n}
	\left( \frac{dT_n}{dE} \right)^2}{[\sum_{n} (d\phi_{n}/dE)]^{2}}.
	\label{rvqpc}
\end{eqnarray}
It is sensitive to the variation with energy of the transmission probability. 
Note that the transmission probability has the form of a Fermi function.
Consequently, the derivative of the transmission probability is 
also proportional to $T_n R_{n}$. The numerator of Eq.~(\ref{rvqpc})
is thus also maximal at the onset of a new channel and vanishes on 
a conductance plateau.

In Fig.~\ref{R_fig} the effective resistance $R(\omega,V)$ is shown 
for four frequencies $\hbar\omega/(eV)=0,0.25,0.5,1$, 
where $V$ is the applied voltage. 
At the highest frequency  $\hbar\omega/(eV)= 1$ 
the resistance $R(\omega,V)$ is completely dominated by the equilibrium 
charge relaxation resistance $R_q$. 
The uppermost curve (d) of Fig.~\ref{R_fig} is nothing but $R_q$ and 
determines the noise due the zero-point equilibrium fluctuations. 
The fluctuations reach a maximum at the onset of 
a new channel since $R_q$ takes its maximum value, $R_q = h/e^{2}$.
At the lowest frequency $\hbar\omega= 0$ the resistance $R(\omega,V)$
is determined by $R_v$. 
The lowermost curve (a) of Fig.~\ref{R_fig} is the nonequilibrium resistance $R_v$. 
It is seen that the non-equilibrium resistance $R_v$ is
very much smaller than $R_q$. We will encounter such a large difference between 
these two resistances also for the chaotic cavity. 
Furthermore $R_v$ exhibits a double peak 
structure: The large peak in the density of states at the threshold 
of a quantum channel nearly suppresses the non-equilibrium noise 
at the channel threshold completely. Two additional 
curves (b and c for $\hbar\omega/(eV)= 0.25$ and  $\hbar\omega/(eV)= 0.5$) describe 
the crossover from $R_v$ to $R_q$.

\section{Quantum chaotic cavity} \label{qd}

\begin{figure}
\narrowtext
\vspace*{1cm}
\epsfysize=6cm
\epsfxsize=7cm
\centerline{\epsffile{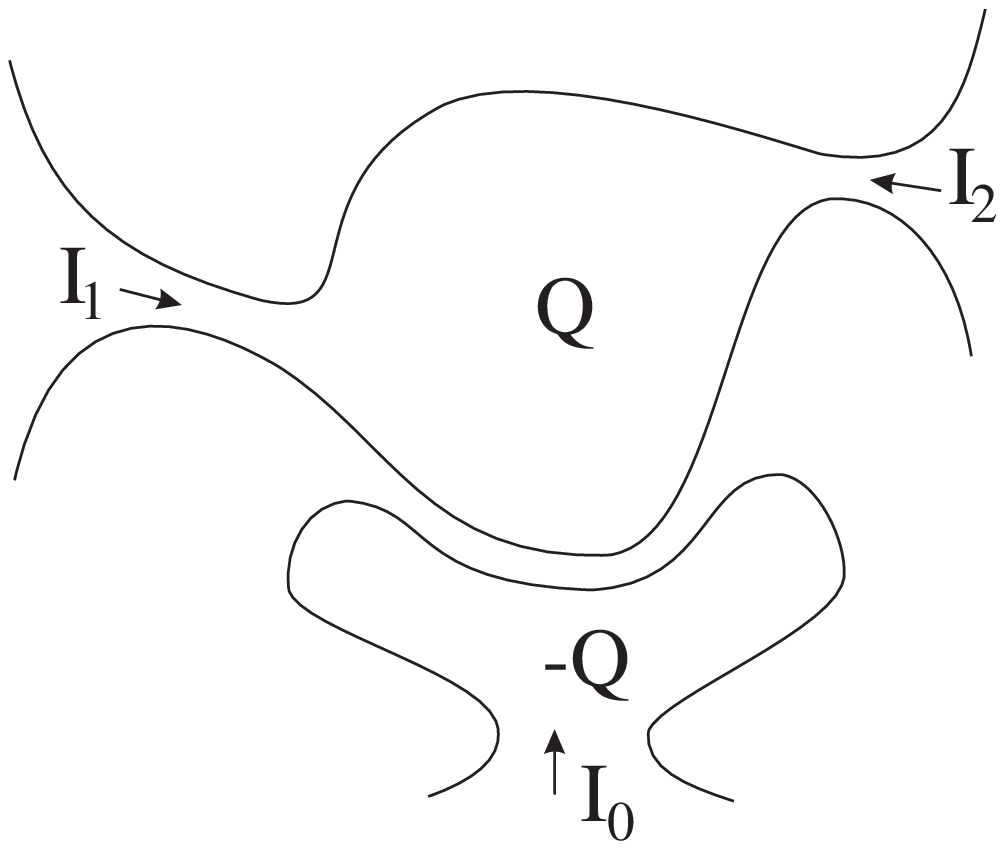}}
\vspace*{1cm}
\caption{ \label{cav}
Quantum dot coupled to two open leads and coupled to a gate.}
\end{figure}

The general theory is now applied to a chaotic 
quantum dot\cite{marcus,chang,chan} with two ideal
single-channel leads and capacitive coupling to a macroscopic 
gate as shown schematically in Fig.~\ref{cav}. 
For such samples, averages lose their meaning and below 
we give the distribution functions of the resistances
which characterize 
the noise induced into the gate contact. 
We compute the statistical distribution of the 
charge relaxation resistance $R_q$ and the resistance $R_v$
from random matrix theory\cite{beenakker}, 
assuming that the classical dynamics of the 
cavity is fully chaotic. We will again consider the 
case $e^{2}/C \gg N^{-1}$ for which the distribution function\cite{note}
of the electrochemical capacitance becomes very sharp.

The distribution of the two-terminal, zero-frequency shot noise
$S = T(1-T)$ in units of $S_{0} =2e (e^{2}/h) |V|$, follows from the distribution 
$P(T) \propto T^{-1+\beta/2}$ of
the dimensionless conductance $T$. The symmetry index $\beta$ equals $1$ or
$2$ depending on whether time-reversal symmetry is present or broken. 
The latter is a result of the uniform distribution of the scattering matrix on
the set of unitary ($\beta=2$) or unitary symmetric ($\beta=1$)
$2\times 2$ matrices\cite{jalabert}. 
Thus one finds for the distribution of $S\in[0,1/4]$ 
\begin{equation}
P(S)= \left\{
\begin{array}{ll}
\frac{\sqrt{1+\sqrt{1-4S}}+\sqrt{1-\sqrt{1-4S}}}{\sqrt{16S(1-4S)}}, &\beta=1, \\
(1/4-S)^{-1/2}, &\beta=2.
\end{array}
\right.
\end{equation}
For the orthogonal ensemble the distribution of the shot 
noise is bimodal and has square root singularities 
at $S= 0$ and $S = 1/4$.
In the unitary ensemble the distribution 
remains finite at zero shot noise and has a square root
singularity only at $S = 1/4$.

\begin{figure}
\narrowtext
\epsfysize=9cm
\epsfxsize=7cm
\centerline{\epsffile{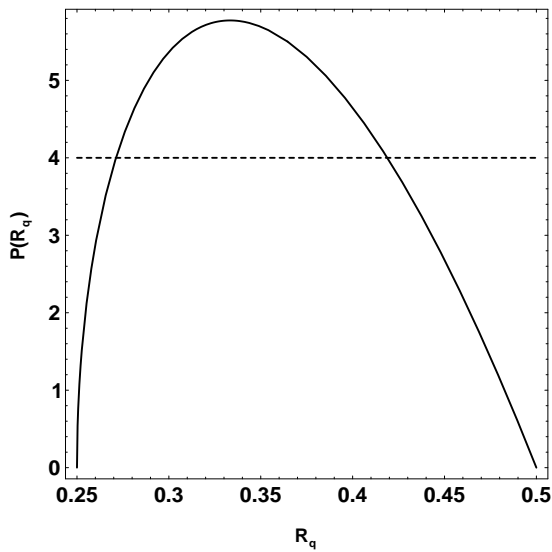}}
\caption{ \label{qdrq}
Distribution of the charge relaxation resistance of a chaotic quantum dot
for the orthogonal ensemble (dashed) and the unitary ensemble (solid line) 
line.}
\end{figure}

In contrast, for the low-frequency spectrum Eq.~(\ref{qfluct}) of the charge
fluctuations, one needs the matrices ${\cal N}_{\alpha\beta}$, which are 
just blocks of the well-known Wigner-Smith delay-time matrix
$\frac{1}{2\pi i}s^\dagger \frac{ds}{dE}$ \cite{wignersmith}. 
The distribution of this matrix has recently been found \cite{brouwer2}.
To compute $P(R_q)$ it is sufficient to know the joint distribution of the
eigenvalues $P(\{q_i\})$, whereas for $P(R_v)$ we also need
that the eigenvectors are distributed uniformly, and independently from
the eigenvalues. Like in Ref.\ \cite{brouwer3} we integrate
over the eigenvalues with an extra weight factor $\sum_i q_i$, which is the
fluctuating density of states. For instance, the distribution of $R_q$
(in units of $h/e^2$) follows from
\begin{equation}
P(R_q)=\int dq_1 dq_2 P(q_1,q_2) (q_1+q_2)
\delta \left( R_q-\frac{q_1^2+q_2^2}{2(q_1+q_2)^2} \right).
\end{equation}
The weight factor appears because, in the limit $e^2/C \gg N^{-1}$, the
ensemble is generated either by {\it uniformly varying} the total charge
(rather than $E_F$) in case the gate voltage is swept, or at {\it constant}
charge (rather than $E_F$) if an other parameter like the magnetic field is
swept. In both cases the average can be replaced by a random matrix average,
provided the density of states is used as a Jacobian \cite{brouwer3}. 
Thus we find the distribution of
the charge relaxation resistance $R_q\in[1/4,1/2]$ 
\begin{equation}
    P(R_q) = \left\{ 
	\begin{array}{ll}
	    4, & \beta=1, \\
	    30(1-2R_q)\sqrt{4R_q-1}, & \beta=2.
	\end{array}
	\right.
\end{equation}
It is shown in Fig.~\ref{qdrq}.

\begin{figure}
\epsfysize=9cm
\epsfxsize=7cm
\centerline{\epsffile{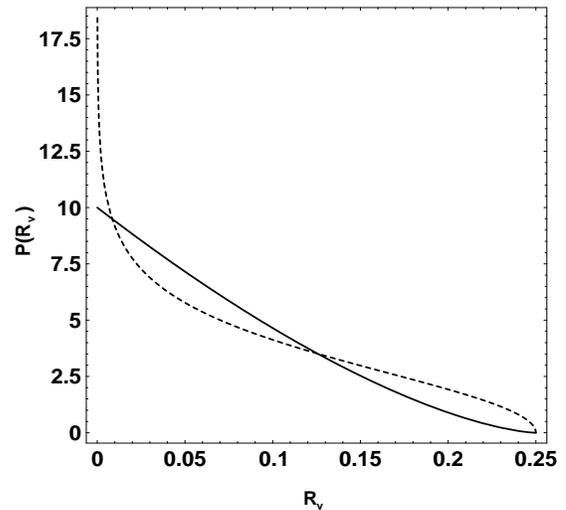}}
\caption{ \label{qdrv}
Distribution of the resistance $R_v$ of a chaotic quantum dot
for the orthogonal ensemble (dashed) and the unitary ensemble (solid line) 
line.}
\end{figure}

For the resistance $R_v$ (also in units of $h/e^2$) 
the distribution is shown in Fig.~\ref{qdrv}.
It is limited to the range
$R_v \in [0, 1/4]$ and given by 
\begin{equation}
P(R_v) = \left\{ 
\begin{array}{ll}
2 \log \left[\frac{1-2R_v+\sqrt{1-4R_v}}{2R_v}\right], & \beta=1, \\
10 (1-4 R_v)^{3/2}, & \beta=2.
\end{array}
\right.
\end{equation}
For the orthogonal ensemble the distribution is singular at $R_v = 0$.
Both distribution functions tend to zero at $R_v = 1/4$.

We see that, as for the quantum point contact, 
the resistance $R_v$ is always smaller than the charge relaxation 
resistance $R_q$. The distributions shown in Figs.\ \ref{qdrq} and \ref{qdrv}
demonstrate that interesting information can be obtained from 
the measurement of frequency-dependent shot noise on chaotic 
quantum dots. 

\section{Discussion} \label{conclusion}

We have investigated the spectrum of the current noise 
induced into the gate of a quantum point contact and 
of a chaotic cavity. 
The current noise spectrum is a direct measure of the charge 
or potential fluctuations of the conductor. 
For this calculation, we have assumed that the external circuit 
exhibits zero impedance for the fluctuations. 
If the impedance of the external circuit 
is not zero, it is necessary to investigate 
also the effect of fluctuating reservoir voltages. 
The fluctuation spectra of a mesoscopic sample embedded in a 
circuit with non-vanishing impedance will then also depend 
on the properties of the external circuit. 

We have treated interactions in random phase approximation. 
Since exchange effects\cite{blanter97,stijn97} play a role, 
a treatment of interactions on the Hartree-Fock level 
is very desirable. The importance to go beyond the 
single parameter potential approximation and to treat 
a continuous potential distribution 
has already been emphasized.
A theory exists already for the low-frequency
fluctuations of a mesoscopic capacitor\cite{buttik96}.

We have found it useful to express the 
noise spectra with the help of resistances,
$R_q$ and $R_v$. The charge relaxation resistance has 
a clear physical meaning since it also determines 
the dissipative, low-frequency admittance of a mesoscopic
conductor\cite{btp93}. The charge relaxation resistance differs from the 
two terminal resistance which one might naively want to use 
to characterize charge relaxation. Whether the resistance $R_v$ 
introduced here will be useful beyond the discussion of noise
properties is not presently apparent. 

The current fluctuations induced into the gate are proportional 
to the square of the electrochemical capacitance of the conductor 
to the gate. The noise will thus be the smaller the more 
effectively the charge on the conductor is screened. 
The strong dependence on interaction of the properties discussed in this 
work are another illustration of the importance of 
screening in the discussion of dynamical effects 
in mesoscopic samples. 

The frequency-dependent noise induced into a nearby gate is a first 
order effect: It is not a small correction to an effect that exists 
already in the zero-frequency limit. 
This lets us hope that experimental detection of this noise 
is possible. From our work it is clear that such 
experiments would greatly contribute to our understanding 
of the dynamics of mesoscopic conductors and the role of interactions. 

\acknowledgments

The work of M.~H.\ P.\ and M.\ B.\ was supported by the Swiss National Science
Foundation. S.~A.~v.\ L.\ was supported by the ``Stichting voor Fundamenteel Onderzoek
der Materie'' (FOM) and by the TMR network ``Dynamics of Nanostructure''.



\end{multicols}

\end{document}